\documentstyle[12pt]{article}

\begin{document}

\begin{titlepage}

\begin{flushright}
RAL-TR-97-070\\
\end{flushright}

\baselineskip 24pt
\begin{center}
{\Large {\bf Standard Model with Duality: Physical Consequences}}\footnote{
Review talk given by the first author at the Krakow Summer School held in 
May-June 1997 at Zapopane, Poland, to appear in Acta Physica Polonica.}\\
\vspace{.3cm}
\baselineskip 16pt
{\large CHAN Hong-Mo}\\
{\it chanhm@v2.rl.ac.uk}\\
{\it Rutherford Appleton Laboratory,}\\
{\it Chilton, Didcot, Oxon, OX11 0QX, United Kingdom}\\
\vspace{.2cm}
{\large TSOU Sheung Tsun}\\
{\it tsou@maths.ox.ac.uk}\\
{\it Mathematical Institute, University of Oxford}\\
{\it 24-29 St. Giles', Oxford, OX1 3LB, United Kingdom}\\
\end{center}

\begin{abstract}
The Dualized Standard Model offers a natural place both to Higgs fields and
to fermion generations with Higgs fields appearing as frame vectors in
internal symmetry space and generation appearing as dual colour.  If they
are assigned those niches, it follows that there are exactly 3 generations
of fermions, and that at the tree-level, only one generation has a mass 
(fermion mass hierarchy) while the CKM matrix is the identity.  However,
loop corrections lift this degeneracy giving nonzero CKM mixing and masses 
to fermions of the two lower generations.  A recent calculation to 1-loop 
level, with just a few parameters, yields a very good fit to the empirical
CKM matrix and sensible values also to the lower generation masses.  In 
addition, predictions are obtained, in areas ranging from low energy 
flavour-changing neutral current decays to extremely high energy cosmic 
rays, which are testable in experiments now being planned.
\end{abstract}

\end{titlepage}

\clearpage

As is well-known, the Standard Model works very well in reproducing 
experimental results.  However, it contains within itself also a number 
of widely-recognized shortcomings.  For instance, at the more fundamental 
level, Higgs fields and fermion generations are introduced into the model 
as phenomenological requirements to fit experimental facts without any 
theoretical reason being given for why they should be there in the first 
place.  At the more practical level, this basic lack of understanding 
is reflected in the large number of empirical parameters.  And even more 
disturbingly, the values of these parameters are seen to reveal some startling 
patterns of the greatest physical significance, and yet no theoretical 
explanation is given for their existence.  For example, there is first 
the so-called fermion mass hierarchy puzzle.  The masses of the $U$-type 
quarks, for example, are quoted in the latest databook \cite{databook} 
as $m_t = 176 \pm 5$ GeV, $m_c = 4.1 - 4.5$ GeV, $m_u = 3 - 8$ MeV, 
dropping by more than two orders of magnitude from generation to generation.  
The mass patterns of the $D$-type quarks and charged leptons are similar, 
although the drop from generation to generation is there a little less 
dramatic.  Then secondly, there is the peculiarity in the mixing pattern,
say for example between the $U$-type and $D$-types quarks as parametrized 
by the CKM matrix \cite{CKM}.  This matrix is found experimentally to be
close to unity but yet differs significantly from it, with off-diagonal 
elements ranging in magnitude from about 22 percent to a few parts permille.
Now these, the fermion masses and CKM matrix elements, are all parameters
on which the properties of our whole physical world crucially depend.  It
would surely be disappointing if we can find no undestanding at all why
they should take the values they take or fall into the noted patterns as
they do.

Answers to these questions are often sought for from beyond the Standard
Model but with, to our minds, no obvious great success.  The difficulty
is that in going beyond the Standard Model, one opens up a wide range of 
freedom for theoretical constructions and so often ends up by putting
in more than one gets out.  In this paper, we wish to describe an approach
we have recently suggested \cite{Chantsou}, in which we attempt to find answers 
for these same questions from within the framework of the Standard Model itself.
This would have at least the advantage of economy and restraint, if it
proves at all possible.  The Standard Model, however, has already been closely
studied, and it is not obvious that one can still find room enough in it
to accommodate the structures of present interest.  On the other hand,
gauge theory is extremely rich in structure, and there are areas in it 
which are still largely unexplored.  

Our new proposal, indeed, is based on a nonabelian generalization of 
electric-magnetic duality which was discovered only recently \cite{Chanftsou}.
When combined with 't~Hooft's famous result \cite{thooft} on confinement 
of 1978, which we shall refer to here as the 't~Hooft Theorem, this generalized
electric-magnetic duality becomes extremely powerful, and when interpreted
in a certain way, leads to the following results.  First, it gives rise
to a new quantum number which shares some properties with the fermion
generation index, and to certain scalar fields which can play the role 
of Higgs fields.  Secondly, if these two abstract quantities occuring 
naturally in the theory are identified respectively with the physical
objects that they resemble, then one predicts straightforwardly (i) that 
there exist 3 and only 3 generations of fermions of each type, (ii) that 
there is a fermion mass hierarchy with one generation of fermions much 
heavier than the other 2, and (iii) that the CKM matrix is close to the 
identity matrix.  All these features are qualitatively as experimentally 
observed.

At the tree-level, the 2 lower generation fermions of each type have
vanishing masses and the CKM matrix equals the identity matrix.  Loop 
corrections, however, lift this tree-level degeneracy and give small 
but nonzero values both to the masses of the 2 lower generation fermions 
and to the off-diagonal CKM matrix elements.  What is more, these 
corrections are found to be calculable perturbatively.  A calculation to
1-loop has already been carried out.  It is found that with just a few 
parameters one is able to obtain a very good fit to the empirical CKM 
matrix as well as sensible values to the fermion masses, a result which
we shall summarize later.

Further, as a consequence to the approach, new Higgs and gauge bosons 
carrying generation indices are predicted at masses in the 100 TeV range.  
Their direct detection is probably out of reach even for LHC, but their
exchange will lead to new effects which may be detectable by some
experiments now being planned.  Of these effects, we shall consider
2, namely (a) flavour-changing neutral current decays and (b) cosmic
ray air showers at ultra-high energies.  As we shall see, interesting 
results are obtained for both.

In this paper, we shall deal only with the general framework and the 
physical consequences.  For the theoretical basis of the approach,
the reader is referred to our companion paper \cite{companion} in the 
same volume.

To set up the framework for our discussion, let us first recall the basic 
tenets of electric-magnetic duality.  As is well-known, electromagnetism 
is dual symmetric in the sense that in addition to the Maxwell potential
$A_\mu(x)$ with which the theory is usually described there exists also
(under certain conditions) a dual potential $\tilde{A}_\mu(x)$ which is
to magnetism what $A_\mu(x)$ is to electricity.  Thus, for example, while
in the description in terms of $A_\mu(x)$, the electric charge $e$ is a 
source and the magnetic charge $\tilde{e}$ is a monopole, so in the 
description in terms of $\tilde{A}_\mu(x)$, the electric charge will
appear as a monopole while the magnetic charge will appear as a source.
Further, since the theory possesses a local $U(1)$ gauge symmetry, it
will possess also a local gauge symmetry under the dual gauge group 
$\tilde{U}(1)$, so that the theory is invariant in all under the doubled
gauge symmetry $U(1) \times \tilde{U}(1)$, where $\tilde{U}(1)$ is the
same group as $U(1)$ but has the opposite parity.  Notice, however, that
although the gauge symmetry is doubled so that there seems to be twice
the usually acknowledged number of gauge degrees of freedom, the number 
of physical degrees of freedom remains the same.  This can be seen as
follows.  In electromagnetism, the dual transform is an operation by the 
usual Hodge star:
\begin{equation}
*F_{\mu\nu}(x) = -\frac{1}{2} \epsilon_{\mu\nu\rho\sigma} F^{\rho\sigma}(x),
\label{Fstar}
\end{equation}
which means in electromagnetism essentially just interchanging $E$ and $H$.
Eq. (\ref{Fstar}) is an algebraic relation giving $*F$ explicitly in terms of 
$F$ so that obviously no new physical degrees of freedom have been introduced.
Nevertheless, with $F$ derivable from a potential $A$, thus:
\begin{equation}
F_{\mu\nu}(x) = \partial_\nu A_\mu(x) - \partial_\mu A_\nu(x)
\label{FinA}
\end{equation}
and $*F$ given by a similar expression in terms of $\tilde{A}$, one sees that
one can perform independent gauge transformations on $A$ and $\tilde{A}$
without changing either the physically measurable quantities $F$ and $*F$
or the dual relation between them.  In spite of the doubled gauge degree
of freedom, therefore, there is only one photon, not two.

Given the importance of nonabelian gauge theories to particle physics, a 
natural question to ask is whether the above dual symmetry known in 
electromagnetism extends to Yang-Mills fields.  If duality is defined
by the Hodge star in (\ref{Fstar}) as in the abelian theory, then the 
answer has long been known to be no \cite{Guyang}.  There is in general 
no guarantee in the nonabelian theory for the existence of the dual 
potential $\tilde{A}_\mu(x)$ having the same relation to $*F_{\mu\nu}(x)$ 
as $A_\mu(x)$ has to $F_{\mu\nu}(x)$.  However, it was shown in 
\cite{Chanftsou} that if one defines the dual transform differently, thus:
\begin{equation}
\omega^{-1} \tilde{E}_\mu \omega = - \frac{2}{\bar{N}} 
   \epsilon_{\mu\nu\rho\sigma} \dot{\eta}^\nu \int \delta \xi ds
   E^\rho \dot{\xi}^\sigma \dot{\xi}^{-2} \delta(\xi-\eta),
\label{Etilde}
\end{equation}
which reduces to (\ref{Fstar}) for the abelian but not for the nonabelian
theory, then the dual symmetry is restored, meaning that there is a dual
potential $\tilde{A}_\mu(x)$ bearing the same relationship to $\tilde{E}_\mu$ 
as the ordinary Yang-Mills potential $A_\mu(x)$ to $E_\mu$.  Here the field
is given in terms of some loop-dependent quantity $E_\mu$ instead of the 
usual local field tensor $F_{\mu\nu}(x)$, and the integral in (\ref{Etilde})
is also a loop integral, so that the generalized dual transform is not
as simple as it looks.  However, for the discussion in this paper of the 
physical consequences of the approach, we shall not need to know much
of the details involved.  In the companion paper \cite{companion}, we shall 
summarize for the theoretically-minded reader the rather intricate arguments 
leading to the above conclusion.  Here, we need to note only in (\ref{Etilde}) 
the quantity $\omega(x)$ which is a local transformation matrix transforming 
from the local frame in which $E_\mu$ is measured to the dual frame in which
$\tilde{E}_\mu$ is measured.  For what follows, this $\omega$ will play
an important role.

As a consequence of the generalized dual symmetry derived in \cite{Chanftsou}, 
one recovers the analogy with the abelian theory desired.  Take for 
example the standard chromodynamics with an $SU(3)$ gauge symmetry.  
In the description in terms of $A_\mu(x)$, a colour electric charge 
$g$ is a source while a colour magnetic charge $\tilde{g}$ is a monopole, 
but in a description in terms of $\tilde{A}_\mu(x)$, $g$ will appear 
as a monopole while $\tilde{g}$ will appear as a source.  And the theory 
will have in all the parity-doubled local gauge symmetry $SU(3) \times 
\widetilde{SU}(3)$, in close parallel to the $U(1) \times \tilde{U}(1)$ 
symmetry of electromagnetism noted above.  Again, however, as in 
electromagnetism, this doubling of the gauge symmetry implies no increase
in the number of physical degrees of freedom.

Next, let us turn to what we call here the 't~Hooft Theorem \cite{thooft}.  
This says that in a theory with gauge symmetry $SU(N)$ (which we call here
generically ``colour''), if colour is confined, then dual colour is Higgsed,
and conversely, if colour is Higgsed, then dual colour is confined.  By 
duality here, however, one means a certain commutation relationship between 
two loop-dependent operators, called by 't~Hooft the order-disorder parameters
$A(C)$ and $B(C)$, the former being the Wilson phase factor:
\begin{equation}
A(C) = {\rm Tr} P \exp\{i g \int_C A_\mu(x) dx^\mu\}.
\label{AC}
\end{equation}
A priori, this need not mean the same thing as the duality discussed in 
the preceding paragraph as defined in (\ref{Etilde}).  However, we were
able to show in a recent paper \cite{Chantsou1} that the operator $B(C)$
constructed in the same way as (\ref{AC}) above, but with instead of 
$A_\mu(x)$ the dual potential $\tilde{A}_\mu(x)$ as given by (\ref{Etilde}),
and instead of $g$ a dual coupling $\tilde{g}$ related to $g$ by the Dirac
quantization condition:
\begin{equation}
g \tilde{g} = 4 \pi,
\label{Diraccond}
\end{equation}
does satisfy the 't~Hooft commutation relation, so that the two usages
of the term duality are in fact interchangeable and that the results can
be combined.

If that is the case, then the combination would be extremely powerful, 
leading to some very interesting consequences.  In particular, suppose 
we apply it to the Standard Model with the gauge symmetry $SU(3) \times SU(2) 
\times U(1)$.  Then according to \cite{Chanftsou}, the theory would also 
have a dual gauge symmetry $\widetilde{SU}(3) \times \widetilde{SU}(2) 
\times \tilde{U}(1)$.  Further, since in the usual interpretation of 
the Standard Model $SU(3)$ colour is confined while the electroweak 
$SU(2)$ symmetry is broken and Higgsed, the 't~Hooft theorem \cite{thooft}
implies that the dual colour symmetry $\widetilde{SU}(3)$ would be broken 
while dual weak isospin, corresponding to the symmetry $\widetilde{SU}(2)$, 
would be confined.  In other words, the roles of the two dual groups would 
be interchanged.

How would these dual symmetries, if they exist, manifest themselves in
the physical world?  Consider first dual colour.  Presumably, as in other
symmetries, particles will form representations of this dual group.  In
particular, we expect that there will be fermions in the fundamental 
representation of $\widetilde{SU}(3)$ forming triplets of dual colour.
However, the symmetry being broken, the members of a triplet will behave
somewhat differently, having for example possibly different masses.  They 
would thus be rather like the members of different generations of any 
particular fermion type, say the $U$- or $D$-type quarks or the charged
leptons or neutrinos.  In other words, dual colour, necessarily broken, would
seem to offer a natural niche for the generation index to fit into.  The
beauty is that if the generation index is assigned that niche, then it 
follows that there will naturally be 3 and only 3 generations of fermions,
a fact which seems to be strongly supported by present experiment.

Would such an assignment work?  To answer this question, we have first to
understand a little about the symmetry-breaking pattern of dual colour.
As for other gauge symmetries, such as weak isospin, we expect 
the dual colour group also to be broken spontaneously via the Higgs
mechanism.  But are there scalar fields around in the theory to play
the role of Higgs fields for breaking this symmetry, or do they have to be 
introduced {\it ad hoc} as in the usual formulation of the electroweak theory?
An interesting feature of the dual framework is that there indeed are scalar
fields occuring naturally in the theory which have the potential for being 
Higgs fields.  Recall the transformation matrix $\omega$ introduced in
the dual transform (\ref{Etilde}) above.  For colour, this is a $3 \times 3$ 
space-time dependent unitary matrix transforming from the colour frame to
the dual colour frame.  Its columns therefore transform as a $3$ of dual
colour, i.e. as the fundamental representation of $\widetilde{SU}(3)$,
while its rows transfrom as a ${\bar 3}$ of colour $SU(3)$.  Under Lorentz
transformations, however, they are space-time scalars.  Moreover, the
matrix $\omega$ being unitary, its rows and columns have unit (nonvanishing)
lengths.  They share therefore many properties that one would want for 
the vacuum expectation values of Higgs fields.  Indeed, if one repeats
the same arguments for the electroweak theory, one finds that they would
work very well as the Higgs fields normally required for symmetry-breaking
in that theory.

At first sight, it might seem rather revolutionary to consider the rows and 
columns of $\omega$ as Higgs fields, but at second look, the move is not so 
new as it appears.  The rows and columns of $\omega$ are basically just the 
frame vectors in internal symmetry space.  Their geometrical significance
is therefore not very different from the vierbeins in General Relativity, and 
in that theory one is used to regarding the vierbeins as dynamical variables.
In making the frame vectors in internal space into Higgs fields, one is in a 
sense copying what is standard procedure in relativity, and at the same
time giving to Higgs fields a geometrical significance which they otherwise
lack.  Conceptually, of course, a geometrical significance for Higgs fields 
would be most welcome in the gauge theory framwork where all the other (gauge
and fermion) fields are known to have deep geometrical significance.

Suppose then we accept this proposal and apply it to $\widetilde{SU}(3)$
of dual colour.  One obtains first 3 dual colour triplets of Higgs fields
and, if one makes what seems the simplest assumption about the dual 
hypercharges they carry, the further result that $\widetilde{SU}(3)$ is 
completely broken with no residual symmetry \cite{Chantsou}.  

Let us denote these Higgs fields by $\tilde{\phi}^{(a)}_{\tilde{a}}$, where 
$(a) = 1, 2, 3$ are just labels for the 3 triplets while $\tilde{a} = 1, 2, 3$ 
denotes the dual colour or the generation of their components.  How would
they couple to the fermions?  We recall that the couplings of Higgs fields
to fermions (Yukawa couplings) are what give us the tree-level fermion mass
matrices.  Now fermions are known to exist in 3 generations so that it will
be natural to assign them in the present scheme to dual colour triplets,
thus: $\psi_{\tilde{a}}, \tilde{a} = 1, 2, 3$.  However, not both the
left- and right-handed fermions can be assigned to dual colour triplets or
otherwise one cannot construct a Yukawa coupling for them.  Taking then 
a cue from the Weinberg-Salam electroweak theory, one proposes to make
the left-handed fermions $\psi_L$ dual colour triplets and the right-handed
fermions $\psi_R$ dual colour singlets.  In that case, one can write a
Yukawa coupling in the form:
\begin{equation}
\sum_{(a)} \sum_{[b]} Y_{[b]} \bar{\psi}^{\tilde{a}}_L 
   \tilde{\phi}^{(a)}_{\tilde{a}} \psi^{[b]}_R.
\label{Yucoup}
\end{equation}

To obtain the tree-level fermion mass matrix, one just substitutes as
usual for the Higgs fields their vacuum expectation values, which we may
take as:
\begin{equation}
\tilde{\phi}^{(1)} = \left( \begin{array}{c} x \\ 0 \\ 0 \end{array} \right);
\tilde{\phi}^{(2)} = \left( \begin{array}{c} 0 \\ y \\ 0 \end{array} \right);
\tilde{\phi}^{(3)} = \left( \begin{array}{c} 0 \\ 0 \\ z \end{array} \right),
\label{phivac}
\end{equation}
giving:
\begin{equation}
m = \left( \begin{array}{ccc} xa & xb & xc \\ ya & yb & yc \\ za & zb & zc
   \end{array} \right) = \left( \begin{array}{c} x \\ y \\ z \end{array} \right)
   (a, b, c).
\label{massmat0}
\end{equation}
with $x, y, z$ real and $a = Y_{[1]}, b = Y_{[2]}, c = Y_{[3]}$ in general
complex.  One notes that the mass matrix is factorizable as indicated.

Now a factorizable mass matrix as that in (\ref{massmat0}) leads to 2
important immediate consequences.  First, as noted already by many others 
but particularly by Fritsch \cite{Fritsch}, a rank 1 matrix such as 
(\ref{massmat0}) has only one nonzero eigenvalue which may be interpreted 
as one generation having a much larger mass than the other two generations 
and hence a reasonable zeroth-order description of the observed fermion 
mass hierarchy mentioned at the beginning.  Second, the first factor 
$(x, y, z)$ in (\ref{massmat0}) depends only on the vacuum expectation
values of the Higgs fields but not on the fermion type, so that the CKM
matrix, which depends only on the first factor and not on the second, will
automatically be the identity matrix at tree-level, again a reasonable
zeroth-order description of the empirical CKM matrix as already mentioned. 
The interesting thing is that both these features which are much desired 
by phenomenology here follow spontaneously as a natural consequence of 
the theoretical framework.

The above scenario, though desirable as a zeroth-order approximation, is
obviously not good enough as a realistic description of nature, for the
2 lower generation fermions, though light compared with the highest
generation, are not actually massless, and the CKM matrix does in fact 
differ considerably from the identity.  Hence, the scheme can only be
considered acceptable if it is capable also of explaining nonzero lower
generation masses and the deviations of the CKM matrix from the identity.
Our contention, to be supported by the results summarized below, is that 
it is indeed capable of doing so via loop corrections.  

Within the scheme, there are of course many different types of possible 
loop diagrams exchanging gluons or dual gluons, Higgses or dual Higgses.  
However, because of the unusual properties built into the scheme, these
loop diagrams share a common property, namely that they leave the mass 
matrix factorizable.  That this is so can be seen as follows.  First, 
ordinary colour gluons and the standard electroweak Higgses do not affect 
the generation (i.e. dual colour) index since they themselves carry no 
dual colour, and therefore corrections due to loops exchanging these bosons
will leave a factorizable mass matrix still factorizable.  Secondly, although
the dual gluons do affect the generation or dual colour index they can couple 
only to the left-handed fermions.  This means that they will only alter
the left-handed factor of the factorized mass matrix but will leave it 
still factorized.  Lastly, the dual colour Higgses both affect the dual
colour index and couple to both left- and right-handed fermions and so
are a potential danger to factorizability, but their couplings to the
fermions, being closely related to the mass matrix, is itself factorizable.
Because of this, it is not difficult to see that loops of dual colour Higgses
will also leave the mass matrix factorizable.  Thus, by examining each type 
of exchanges in turn, one easily convinces oneself that no 1-loop diagram 
of any type will modify the factorizable nature of the mass matrix.  Indeed, 
we are of the opinion, though cannot claim to have rigorously demonstrated 
the fact, that even higher loops of any order will still leave the 
factorizability of the mass matrix intact.

Although the mass matrix remains factorized under loop corrections, this
does not mean necessarily that the lower generation masses must remain zero
or that the CKM matrix must remain the identity.  For the CKM matrix, this 
is readily seen.  For example, the dual gluon loop, as explained above, 
though leaving factorizability intact, modifies the left-hand factor of 
the mass matrix, and this modification can depend on whether the fermion 
addressed is the $U$-type or the $D$-type quark.  Hence, the CKM matrix 
which basically measures the relative orientation of these left-hand 
factors of respectively the $U$-type and $D$-type quarks need no longer 
be the identity when the dual gluon loop correction is taken into account.

That the lower generation fermion masses also need not remain zero after
loop corrections is not so obvious and is in fact quite intriguing.  Given
that the mass matrix is still factorizable after loop corrections and hence
still of rank 1, it follows that it has always only one nonzero eigenvalue.
However, it is not obvious that the 2 remaining zero eigenvalues ought to be 
interpreted as the masses of the 2 lower generations.  The point is that, 
though still factorizable, the mass matrix can be rotated by the loop 
corrections and this rotation depends on the renormalization scale.  And 
once a mass matrix has an orientation which is scale-dependent, it is not 
so obvious what ought to be defined as the masses and the state vectors of 
the physical states.  Indeed, we believe that this question would have 
arisen already in the usual (i.e. nondualized) standard model, had the effect 
there not been so negligibly small.  The ambiguity comes about as follows.
Consider first a mass matrix with a scale-independent orientation.  Then 
once it is diagonalized at some scale it will remain diagonal at any other 
scale.  And if it is a hermitian matrix, then its eigenvectors will be 
orthogonal as physical state vectors ought to be.  It would then be 
appropriate, as is usually done, to define the mass of each physical 
state as the appropriate eigenvalue evaluated at the scale equal to its 
value, as one would as if only one state is involved.  However, if the 
matrix rotates as the scale changes, then the eigenvector of one eigenvalue 
at the scale equal to its value will not usually be orthogonal to the 
eigenvetor of another eigenvalue evaluated at the other scale equal 
to its own value.  Hence, these two vectors can no longer be associated 
with two independent physical states.  In fact, we do not know a valid 
criterion for defining the masses and physical states from a general 
mass matrix which rotates with changing scales.

However, for the special case we have here of a factorizable mass matrix,
it is possible to define the masses and the states in such a way that each
mass is evaluated at the scale equal to its value and still have all the 
physical state vectors mutually orthogonal.  Let us illustrate the problem
with the $U$-type quarks.  At any scale, the factorizable mass matrix has 
only one nonzero eigenvalue.  If one evaluates the mass matrix at the scale
equal to this eigenvalue, then one can unambiguously define this value as 
the mass of the top quark, and the corresponding eigenvector as the physical
top state vector.  The other 2 eigenvalues at this scale are zero but they 
ought not for consistency to be identified as the masses of the lower states 
$c$ and $u$ for they would be evaluated at the wrong scale.  Also, at this
stage, we do not know which are the physical state vectors corresponding to
respectively the $c$ and $u$ states.  However, we do know that the state 
vectors of $c$ and $u$, being by definition orthogonal to the state vector 
of the top, have to lie in the subspace spanned by the zero eigenvectors 
of the mass matrix evaluated at this the top-mass  scale.  Suppose now we 
run the scale to a lower energy.  Since the mass matrix rotates with changing 
scales, the 2 zero eigenvectors at the top mass will no longer be zero
eigenvectors at the new scale, so that the $2 \times 2$ mass submatrix in 
the subspace spanned by the 2 originally zero eigenvectors need no longer be 
zero.  However, it will still be factorizable and of rank 1, and has therefore 
again only one nonzero eigenvalue, a situation exactly the same as that
we started with for the full mass matrix.  For logical consistency, therefore, 
one should again evaluate this nonzero eigenvalue of the submatrix at a 
scale equal to its value and define this value as the mass of the second 
generation, namely that of the charm quark $c$.  It follows also that the 
corresponding eigenvector at this scale should be defined as the $c$ 
physical state vector which, by definition, will be automatically orthogonal 
to the top state vector already defined.  Having then defined both the
$t$ and $c$ physical state vectors, one can unambiguously define (up
to a sign) the physical $u$ state vector as the vector orthogonal to 
both.  Furthermore, by repeated the procedure once more running the scale 
even lower in energy, one can clearly define also the $u$ mass.  In this 
way, all masses and state vectors are uniquely defined, each mass is 
evaluated at the scale equal to its value, and the 3 physical state vectors 
are mutually orthogonal, as is appropriate.

Though described above only in words, this procedure for evaluating the
the CKM matrix and the lower generation fermion masses is not merely a 
theoretical prescription but one that can be put into actual practice.
Indeed, a calculation in this direction to 1-loop level has already been 
done \cite{Bordesetal}.  The calculation being somewhat complicated and 
containing a number of quite intriguing details, we have space here only 
to give a bare outline of the main steps involved and to summarize the 
main results.

First, among the many 1-loop diagrams calculated, some are found to be 
large and not calculable perturbatively.  They affect, however, only the 
normalization of the mass matrix, not its orientation which, as explained 
above was of the most interest.  It pays, therefore, at present to abandon 
the calculation of the normalization and focus one's attention just on the 
orientation.  This reduction in objective then removes the necessity for 
evaluating several of the diagrams.  Secondly, on putting in the estimate 
for the dual gluon mass obtained from the stringent empirical bounds on 
flavour-changing neutral current decays, one finds that most of the remaining 
diagrams affect the orientation of the mass matrix only to a negligible 
degree, leaving in the end only one diagram that really matters, namely
the dual colour Higgs loop.  Thirdly, by a rather fortunate accident, the
effect of this last diagram is to a very good approximation independent
of some parameters, such as the Higgs boson masses, on which it formally 
depends.  Fourthly, by adjusting, among the remaining parameters, the 
strength of the Yukawa coupling $\rho$, one finds one can indeed account 
for the fermion mass of the second generation as a `leakage' from the highest 
by the procedure detailed above.  Then fixing the mass scales and Yukawa 
couplings, one each for each fermion type, by fitting the masses of the 
two higher generations, one is finally left with only 2 real parameters, namely 
the 2 ratios between the 3 vacuum expectation values $x, y, z$ of the dual 
colour Higgs fields, with which to evaluate the CKM matrix and all the 
fermion masses of the lowest generation.

Emphasis was put on fitting the CKM matrix, which depends only on the 
orientation, rather than on the lowest generation masses which depend also 
on the as yet incalculable normalization of the mass matrices.  The 
following is a sample of the sort of fits obtained:
\begin{equation}
|V| = \left( \begin{array}{ccc} 
              0.9752 & 0.2215 & 0.0048 \\
              0.2211 & 0.9744 & 0.0401 \\
              0.0136 & 0.0381 & 0.9992 
              \end{array} \right),
\label{CKMout}
\end{equation}
which is to be compared with the following experimental values entered in 
\cite{databook}:
\begin{equation}
|V| = \left( \begin{array}{lll}
      0.9745 - 0.9757 & 0.219 - 0.224 & 0.002 - 0.005 \\
      0.218 - 0.224 & 0.9736 - 0.9750 & 0.036 - 0.046 \\
      0.004 - 0.014 & 0.034 - 0.046 & 0.9989 - 0.9993 \end{array} \right).
\label{CKMexp}
\end{equation}
The agreement is seen to be good.  This we find encouraging since it is 
not at all obvious that the CKM matrix can be so fitted.  In particular
we note the large value of $V_{cd}$ and $V_{us}$, i.e. the Cabibbo angle,
compared with the other elements.  This comes about directly from the
special way described above of how the lower generation states are defined
by `running' and can thus be considered as some confirmation of its validity.  
We note, however, that all CKM matrix elements in the calculation are real, 
so that at least at the 1-loop level we have worked with so far, there is 
no possibility of a CP-violating phase.

There are two rather astounding features common to all the fits we have 
found thus far:  (a) The approximate equality, to a few percent accuracy,
of all the Yukawa coupling strengths $\rho$ for the 3 fermion types that 
we have fitted, namely the $U$- and $D$-type quarks and the charged leptons.  
(b) The proximity, to within around 1 part in ten thousand, of the 
normalized vector $(x, y, z)$ representing the vacuum expectation 
values of the dual colour Higgs fields to one of its fixed points 
$(1,0,0)$.  These seem to us possibly indicative of a deeper symmetry
that we do not yet understand.  In particular, the fitted values of the 
$\rho$'s are so close that one could easily obtain as good a fit as the
best by requiring all $\rho$'s to be identical.  In other words, had we
known a theoretical reason why the $\rho$'s should be the same, we could
have fitted very well all CKM matrix elements and all masses of the second 
generation with only 3 parameters, namely one common $\rho$ and the 2 
ratios of $(x, y, z)$.

Further, an attempt may also be made to estimate the fermion masses of 
the lowest generation by the method outlined above.  However, in contrast
to the calculation of the CKM matrix which depends only on the orientation
of the mass matrix, the estimates for the lowest generation fermion masses 
depend also on the change with scales of the normalization, not only of
the mass matrix but also of the Yukawa coupling strength $\rho$.  If we
naively just assume that both the normalizations of the mass matrix and
$\rho$ are constants independent of scale changes, then, using the same
parameters (\ref{CKMout}) as those determined above in fitting the CKM 
matrix, one obtains:
\begin{equation}
m_u = 235 \ {\rm MeV}, \ \ m_d = 17 \ {\rm MeV}, \ \ m_e = 7 \ {\rm MeV},
\label{mudeout}
\end{equation}
which are to be compared with the experimental values quoted in \cite{databook}:
\begin{equation}
m_u = 2 - 8 \ {\rm MeV},\ \ m_d = 5 - 15 \ {\rm MeV},\ \ m_e = 0.5 \ {\rm MeV}.
\label{mudeexp}
\end{equation}
Considering that in obtaining the values in (\ref{mudeout}) scale-dependences 
of normalizations have been neglected over several orders of magnitude in 
energy, we regard the estimates as quite sensible except perhaps for the 
$u$, for which a change in scale over 4 decades of energy is involved.

Suppose we tentatively accept these results as reasonable confirmation that
the dualized standard model is capable of explaining the mass and mixing 
patterns of fermions, our next question should be whether the scheme leads 
also to new predictions which can be tested against experiment.  The most 
obvious to examine first is the predicted existence of a new batch of 
particles, namely the dual colour gauge and Higgs bosons.  At first sight, 
it appeared that the calculation of 1-loop effects summarized above might 
give an estimate of these bosons' masses, but unfortunately for this purpose, 
though fortunately for the calculation itself as already noted, the result 
is to a good approximation independent of these mass parameters.  On the
other hand, the stringent experimental bounds on flavour-changing neutral 
current decay give a lower limit for the dual gluon mass of around several 
100 TeV, corresponding to a lower limit on the dual Higgs mass of several 
10 TeV.  If these limits are accepted, then it is unlikely that these 
particles can be produced even by the LHC.  However, the exchange of these 
particles can lead to effects detectable in experiment.  Flavour-changing 
neutral current decays, as already noted, are one such example.

Another prediction of this sort, which at first sight looks quite alarming,
is that of a strong interaction for neutrinos at high energy.  This arises
as follows.  Identifying generation with dual colour implies that neutrinos
also carry dual colour and hence will interact via dual gluon exchange.
Now the coupling of the dual gluon is related to the coupling of the gluon
by the Dirac quantization condition (\ref{Diraccond}).  Substituting the 
experimental value of $\alpha_s = (g^2/4\pi) \sim 0.120$ gives a value 
for $\tilde{g}$ of order 10, which is very large.  There is thus predicted 
a very strong interaction between neutrinos due to the dual gluon exchange.  
However, because the dual gluon is very heavy, this interaction will not 
be effective at energies available to present or near future laboratories.  
To search for this effect we shall have to look to cosmic rays.

Now it so happens that there is indeed a long-term puzzle in cosmic 
ray physics which seems explainable as a manifestation of this
phenomenon.  Over the last 30 years or so, a small number of very high 
energy air shower events with primary energy $E > 10^{20}$ eV have been
observed \cite{Baratav}.  They are a mystery because in theory they should 
not exist.  High energy air showers are thought to be mostly due to protons, 
but protons at an energy beyond $5 \times 10^{19}$ eV will quickly lose 
it via its interaction with the 2.7 K microwave background.  Indeed, it was 
shown by Greisen \cite{Greisen}, and by Zatsepin and Kuz'min \cite{Zatsemin},
that protons with energy in excess of that cut-off cannot reach us from
a distance of more than 50 Mpc.  At the same time, protons at such energies
will hardly be deflected by the magnetic fields either in the galaxy or in
intergalactic space so that it should be easy to identify any candiate 
sources within a radius of 50 Mpc.  However, searches along the direction of 
the observed events fail to reveal any likely source within that sort of 
distance.  The conclusion would seem thus to be either that there are some 
rather exotic sources nearby without us knowing about them or that these 
air showers are not due to protons at all but to some other particles.

An interesting possibility for the present scheme is that they are due to
neutrinos having acquired strong interactions at high energy as predicted
above.  This explanation not only seems feasible but appears even capable
of overcoming several difficulties plaguing the proton explanation.  First, 
neutrinos, being neutral in charge, would not interact with the 2.7 K 
microwave background as the protons would and can therefore reach us with 
energy above the Greisen-Zatsepin-Kuz'min cut-off even if they have 
originated from a distant source much beyond 50 Mpc.  Secondly, since they 
are strongly interacting at high energy, they can be produced copiously in 
high energy collisions, say of protons in an active galactic nucleus but, 
in contrast to protons, can escape from the strong radiation fields which 
are thought to surround active galactic nuclei.  Thirdly, when they arrive 
on earth, their strong interactions with the air nuclei together with the 
fact that dual gluons and gluons, as mentioned already in the beginning, 
represent basically the same physical degree of freedom \cite{Bordesetal1},
can give them a sufficiently large cross section to initiate air showers as
observed.  (On this point, we disagree with the conclusion of a recent paper 
by Burdman, Halzen and Gandhi \cite{Burdzendhi} which claims the opposite.
See e.g. also \cite{Bordesetal2}.)  Indeed, working in this direction, it 
is even possible to make a rough estimate for the high energy neutrino-air 
nucleus cross section and hence suggest direct experimental tests for the 
hypothesis that air showers above the Greisen-Zatsepin-Kuz'min cut-off are 
due to neutrinos rather than protons \cite{Bordesetal1}.  It seems thus 
that the prediction of a strong interaction for neutrinos at high energy 
not only may not prove to be an embarassment but may even help to resolve 
a long-term puzzle in cosmic rays physics.

A neutrino at $10^{20}$ eV primary energy impinging on a proton at rest
in the atmosphere corresponds to a CM energy of 400 TeV, which we note
is just above the lower bound imposed on the dual gluon mass by the
present experimental bounds on flavour-changing neutral current decays.
If we take seriously the proposal in the preceding paragraph, it would
appear then that the mere existence of these extremely high energy air showers
would imply also an upper bound on the mass of the dual gluon.  That being
the case, we could turn the argument around and use this estimate for the
dual gluon mass to predict the branching ratios of various flavour-changing
neutral decays which can be tested against experiment.  An attempt in that
direction has already been made \cite{Bordesetal1}, which gives branching
ratios seemingly within reach of some future experiment now being planned.
These predictions can be further sharpened \cite{Bordesetal3} using the 
recent results from the CKM matrix calculation \cite{Bordesetal} described
above.

So far, we have dealt in this review only with the dual colour symmetry
$\widetilde{SU}(3)$ and its interpretation as generation.  However, as
already mentioned above, within the Dualized Standard Model framework, 
there is also a dual weak isospin symmetry $\widetilde{SU}(2)$ which is
confined and might lead to further physical consequences.  Analogy with 
colour confinement would suggest that at low energy, only $\widetilde{SU}(2)$
singlets can exist which are analogues to hadrons of $SU(3)$ colour, but
deep inelastic experiments at high energy could reveal a substructure to
these $\widetilde{SU}(2)$ singlet states analogous to the parton substructure 
of hadrons in colour $SU(3)$.  However, an estimate of the gauge coupling 
$\tilde{g}_2$ of $\widetilde{SU}(2)$ via the Dirac quantization condition 
(\ref{Diraccond}) from the empirical value of the ordinary weak isospin 
coupling $g_2$ gives a value many times larger than the coupling $g_3$ 
responsible for colour confinement, suggesting thus that the energy scale 
required to reveal the $\widetilde{SU}(2)$ substructure may be much higher
than that required for the $SU(3)$ case.  This might mean therefore that
deep inelastic experiments, provided that the energy is high enough, could 
reveal a substructure to what we believe at present to be elementary 
particles, such as quarks and leptons, or the Higgs and gauge bosons.

In any case, the Dualized Standard Model seems not only to offer tentatively
viable explanations to some long-term puzzles in particle as well as in 
astroparticle physics, but can give rise to some interesting new predictions 
which may one day be testable against experiment.

We shall conclude this brief review by a general suggestion on terminology.  In 
the literature, the term `generation' has been used interchangeably with the 
term `family', with perhaps a slight preference for the latter because, in 
the words of Jarlskog \cite{Jarlskog}, there was no known ``mother-daughter
relationship between the copies''.  However, if the explanation for
fermion masses offered above by the Dualized Standard Model is accepted,
then there is now a known mother-daughter relationship between the 3 copies, 
with the higher mass copies indeed giving birth, in a sense, to the masses 
of the lower copies, via the mechanism due to the rotating mass matrix.  
It would seem thus that the term `generation' is most appropriate.  
We suggest therefore that one keeps the term `generation' in the sense it 
has been used throughout this paper, but adopt the term `family' to denote, 
in accordance with the biological usage of the term, the collection of 
members related by this ``mother-daughter'' relationship.  In other words, 
the fermion types would then be labelled as the $U$-family, the $D$-family 
etc., while the $t, c$ and $u$ would be labelled as the members of the 1st, 
2nd, and 3rd generations of the $U$-family.

We wish to thank Jos\'e Bordes, Jacqueline Faridani and Jakov Pfaudler,
our collaboration with whom has generated most of the work reported in
this review as well as a lot of pleasure while doing it.

\end{document}